\documentstyle[aps,epsf,preprint]{revtex}
\tightenlines
\begin{document}
\draft
\title{Proposed Search for \lowercase{\boldmath{$a^0_0(980)-f_0(980)$}}
Mixing in Polarization Phenomena\\[.5cm]}
\author{N.N. Achasov \thanks{Email: achasov@math.nsc.ru}\,\, and
\  G.N. Shestakov \thanks{Email: shestako@math.nsc.ru}\\[.8cm]}
\address{Laboratory of Theoretical Physics,
Sobolev Institute for Mathematics,\\ Novosibirsk, 630090, Russia\\[1cm]}
%\date{}
\maketitle
\begin{abstract}

The $K^+$ and $K^0$ meson mass difference induces the mixing of
the $a^0_0(980)$ and $f_0(980)$ resonances, the amplitude of
which, between the $K^+K^-$ and $K^0\bar K^0$ thresholds, is large
in magnitude, of the order of
$\,m_{K}\sqrt{m^2_{K^0}-m^2_{K^+}}\approx\sqrt\alpha \,m^2_K$\,,
and possesses the phase sharply varying by about 90$^\circ$. We
suggest performing the polarized target experiments on the
reaction $\pi^-p\to\eta\pi^0n$ at high energy in which the fact of
the existence of $a^0_0(980)-f_0(980)$ mixing can be unambiguously
and very easily established through the presence of a strong jump
in the azimuthal asymmetry of the $\eta\pi^0$ $S$ wave production
cross section near the $K\bar K$ thresholds. The presented
estimates of the polarization effect to be expected in experiment
are to a great extent model independent.
\end{abstract}
\vspace*{0.7cm} 

\newpage

Study of the nature of light scalar resonances has become a
central problem of non-perturbative QCD. The point is that the
elucidation of their nature is important for understanding both
the confinement physics and the chiral symmetry realization way in
the low-energy region, i.e., the main consequences of QCD in the
hadron world. The nontrivial nature of well established lightest
scalar resonances is no longer denied practically anybody. In
particular, there exist numerous evidences in favor of the
four-quark ($q^2\bar q^2$) structure of these states, see, for
example, Ref. [1] and references therein. In this Letter, we
propose a new method of the investigation of the $a_0(980)$ and
$f_0(980)$ resonances with use of the polarization phenomena
closely related to the $a_0^0(980)-f_0(980)$ mixing effect that
carries important information on their nature, in particular,
concerning their coupling to the $K\bar K$ channels.

The mixing between the $a_0^0(980)$ and $f_0(980)$ resonances was
discovered theoretically as a threshold phenomenon in the late 70s
[2]. Recently interest in the $a_0^0(980)-f_0(980)$ mixing was
renewed, and its possible manifestations in various reactions are
intensively discussed in the literature [3--16]. For example, in
Ref. [7] it was suggested that the data on the centrally produced
$a_0^0(980)$ resonance in the reaction $pp\to p_s(\eta\pi^0)p_f$,
in principle, can be interpreted in favor of the existence of
$a_0^0(980)-f _0(980)$ mixing. In Ref. [10] it was noted that
within the experimental errors and the model uncertainty in the $f
_0(980)$ production cross section the result obtained in Ref. [7]
made not contradict to the predictions done in Ref. [2]. However,
the experimental confirmation of such a scenario requires
measuring the reaction $pp\to p_s(\eta\pi^0)p_f$ at a much higher
energy to exclude a possible effect of the secondary Regge
trajectories, for which the $\eta\pi^0$ production is not
forbidden by $G$ parity. This Letter presents a qualitatively new
proposal concerning a search for the $a_0^0(980)-f_0(980)$ mixing
effect. We propose the polarized target experiments on the
reaction $\pi^-p\to\eta\pi^0n$ at high energy in which the fact of
the existence of $a^0_0(980)-f_0(980)$ mixing can be unambiguously
and very easily established through the presence of a strong jump
in the azimuthal asymmetry of the $S$-wave $\eta\pi^0$ production
cross section near the $K\bar K$ thresholds.

Owing to parity conservation, the differential cross section of
the reaction $\pi^-p\to(\eta \pi^0)_S\,n$ on a polarized proton
target at fixed incident pion laboratory momentum, $P^{\pi^-}_{lab
}$, has the form \begin{equation} d^3\sigma/dtdmd\psi=[\,d
^2\sigma/dtdm+I(t,m)\,P\cos\psi\,]/2\pi\,,\end{equation} where
$(\eta\pi^0)_S$ denotes a $\eta\pi^0$ system with the relative
orbital angular momentum $L=0$, $t$ is the four-momentum transfer
squared from the incident $\pi^-$ meson to the outgoing $\eta
\pi^0$ system, $m$ is the $\eta\pi^0$ invariant mass, $\psi$ is
the angle between the normal to the reaction plain, formed by the
momenta of the $\pi^-$ and $\eta\pi^0$ system, and the transverse
(to the $\pi^-$ beam axis) polarization of the protons, $P$ is a
degree of this polarization,
$d^2\sigma/dtdm=|M_{++}|^2+|M_{+-}|^2$ is the unpolarized
differential cross section, $M_{+-}$ and $M_{++}$ are the
$s$-channel helicity amplitudes with and without nucleon helicity
flip, $I(t,m)=2\,\mbox{Im}(M_{++}M^*_{+-})$ describes the
interference contribution responsible for the azimuthal (or spin)
asymmetry of the cross section. In terms of the directly
measurable quantities $I(t,m)$ and $d^2\sigma/dtdm$, one can also
define the dimensionless normalized asymmetry
$A(t,m)=I(t,m)/[d^2\sigma/dt dm]$, $\,-1\leq A(t,m)\leq1$ [17].
Here we are interested in the region of $m\approx1$ GeV. The
available data from unpolarized target experiments on the reaction
$\pi^-p\to\eta\pi^0n$, performed with $P^{\pi^-}_{lab}$ of 18.3
GeV at BNL [3,18,19], 38 GeV at IHEP [20,21], 32 GeV at IHEP [21],
and 100 GeV at CERN [21], show that the $(\eta\pi^0)_S$ mass
spectrum in this region of $m$ is dominated by the production of
the $a_0^0(980)$ resonance, $\pi^-p\to
a_0^0(980)n\to(\eta\pi^0)_S\,n$.

From the $G$-parity conservation it follows that at high energies
and small $-t$ the amplitudes $M_{+-}$ and $M_{++}$ are defined by
the $t$-channel exchanges with quantum numbers of the $b_1$ and
$\rho_2$ Regge poles, respectively [4] (hereinafter they are
denoted by $M^{b_1}_{+-}$ and $M^{\rho_2}_{++}$). In addition, the
possibility of the $\pi$ Regge pole exchange in the reaction
$\pi^-p\to(\eta\pi^0)_S\,n$ arises by virtue of the process
$\pi^-p\to f_0(908)n\to a^0_0(980)n\to(\eta\pi^0)_S\,n$ stipulated
by the $G$-parity violating $a^0_0(980)-f_0(980)$ mixing [2,4,22].
As is well known, the amplitude of the $\pi$ exchange is large in
the low $-t$ region. Moreover, the modulus and the phase of the
$a^0_0(980)-f_0(980)$ transition amplitude both change
dramatically as functions of $m$ near the $K\bar K$ thresholds. As
we shall see, all of these features lead in the reaction $\pi^-p
\to(\eta\pi^0)_S\,n$ to rather impressive consequences, which can
be easily revealed in polarized target experiments because of
measuring the interference between the $\rho_2$ and $\pi$ exchange
amplitudes.

Let us now turn to the quantitative estimates of the expected
polarization effect. As for the $G$-parity violating $\pi$
exchange amplitude $M^\pi_{+-}$, essentially all is known,
including its absolute normalization [2,4,23]. This amplitude can
be written as follows:
\begin{equation}
M^{\pi}_{+-}=e^{-i\pi\alpha_{\pi}(t)/2}\,\frac{\sqrt{-t}}{t-m^2_\pi}\,
e^{\Lambda_{\pi}(t-m^2_\pi)/2}\,a_\pi\,e^{i\delta_B(m)}\,G_{a_0f_0}(m)\,[2m^2
\Gamma_{a_0\eta\pi^0}(m)/\pi]^{1/2}\,,\end{equation} where
$\alpha_\pi(t)=\alpha_\pi(0)+\alpha'_\pi
t\approx0.8(t-m^2_\pi)/\mbox{GeV}^2$ is the $\pi$ Regge pole
trajectory, $\,a_\pi=g_{\pi NN}\,g_{f_0\pi^+\pi^-}/\sqrt{8\pi}s$,
$\,g^2_{\pi NN}/4\pi\approx14.3$, $\,g_{f_0\pi^+\pi^-}$ is the
$f_0(980)$ coupling constant to the $\pi^+\pi^-$ channel,
$\,s\approx2m_p P^{\pi^-}_{lab}$,
$\,\Lambda_\pi/2=\Lambda^0_\pi/2+\alpha'_\pi\ln(s/s_0)$ is the
residue slope, $s_0=1$\,GeV$^2$, $\,\delta_B(m)$ is a smooth and
large phase (of about 90$^\circ$ for $m\approx1$ GeV) of the
elastic background accompanying the $f_0(980)$ resonance in the
$S$-wave reaction $\pi\pi\to\pi\pi$ in the channel with isospin
$I=0$ [2,23], $\,G_{a_0f_0}(m)=\Pi_{a_0f_0}(m)/[D_{a_0}(m)
D_{f_0}(m)-\Pi^2_{a_0f_0}(m)]$, $\,\Pi_{a_0f_0}(m)$ is the
nondiagonal matrix element of the polarization operator describing
the $a^0_0(980)-f_0(980)$ transition amplitude [2], $1/D_r(m)$ is
the propagator of an unmixed resonance $r$ with a mass $m_r$,
$\,D_r(m)=m^2_r-m^2+\sum
_{ab}[\mbox{Re}\Pi^{ab}_r(m_{f_0})-\Pi^{ab}_r(m)]$,
$\,r=[a_0(980),f_0(980)]$, $\,ab =(\eta\pi^0,K^+K^-,K^0\bar K^0)$
for $r=a_0(980)$, and $\,ab=(\pi^+\pi^-,\pi^0\pi^0,K^+K^-,K^0\bar
K^0)$ for $r=f_0(980)$, $\,\Pi^{ab}_r(m)$ is the diagonal matrix
element of the polarization operator for the resonance $r$
corresponding to the $ab$ intermediate state contribution [23],
$\,\Gamma_{rab}(m)=\mbox{Im}[\Pi^{ab}_r(m)]/m
=g^2_{rab}\,\rho_{ab}(m)/16\pi m$ is the width of the $r\to ab$
decay, $g_{rab}$ is the coupling constant of $r$ to the $ab$
channel, $\rho_{ab}(m)=[(m^2-m^2_+)(m^2-m^2_-)] ^{1/2}/m^2$, and
$m_\pm=m_a\pm m_b$. The $a^0_0(980)-f_0(980)$ transition amplitude
$\Pi_{a_0f_0}(m)$ must be determined to a considerable extend by
the $K^+K^-$ and $K^0\bar K^0$ intermediate states because of the
proximity of the $a ^0_0(980)$ and $f_0(980)$ resonances to the
$K\bar K$ thresholds and their strong coupling to the  $K\bar K$
channels. The sum of the one-loop diagrams $f_0(980)\to K^+K^-\to
a^0_0(9 80)$ and $f_0(980)\to K^0\bar K^0\to a^0_0(980)$, with
isotopic symmetry for coupling constants, gives [2]
$$\Pi_{a_0f_0}(m)=\frac{g_{a_0K^+K^-}g_{f_0K^+K^-
}}{16\pi}\Biggl[\,i\,\Bigl(\rho_{K^+K^-}(m)-\rho_{K^0\bar
K^0}(m)\Bigr)$$
\begin{equation}\left.-\,\frac{\rho_{K^+K^-}(m)}{\pi}\ln\frac{1+\rho_{K^+K^-}(m)}
{1-\rho_{K^+K^-}(m)}+\frac{\rho_{K^0 \bar
K^0}(m)}{\pi}\ln\frac{1+\rho_{K^0 \bar K^0}(m)}{1-\rho_{K^0\bar
K^0}(m)}\,\right]\,,\end{equation} where $m\geq2m_{K^0}$; in the
region $0\leq m\leq2m_K$, $\rho_{K\bar K}(m)$ should be replaced
by $i|\rho_{K\bar K}(m)|$. The ``resonancelike"\ behavior of the
modulus and the phase of $\Pi_{a_0f_0}(m)$ is clearly illustrated
in Figs. 1(a) and 1(b). Note that in the 8-MeV-wide region between
the $K^+K^-$ and $K^0\bar K^0$ thresholds
$|\Pi_{a_0f_0}(m)|\approx|g_{a_0K^+K^-}g_{f_0K^+K^-}/16\pi|
[(m_{K^0}^2-m_{K^+}^2)/m_{K^0}^2]^{1/2}\approx0.1265
|g_{a_0K^+K^-}g_{f_0K^+K^-}/16\pi|$, that is, of the order of
$m_K\sqrt{m^2_{K^0}-m^2_{K^+}}\approx\sqrt\alpha\,m^2_K$ [2]. From
Eqs. (2) and (3) it follows also that the contribution of $M^\pi
_{+-}$ to $\,d^2\sigma/dtdm$, in this mass region, is controlled
mainly by the production of the ratios $R_1=g^2_{f_0K^+K
^-}/g^2_{f_0\pi^+\pi^-}$ and
$R_2=g^2_{a_0K^+K^-}/g^2_{a_0\eta\pi^0}$, i.e.,
$|M^\pi_{+-}|^2\propto\sigma(\pi^+\pi^-\to\eta\pi^0)\propto
R_1R_2$.

Within the Regge pole model the amplitudes $M^{b_1}_{+-}$ and $M^{
\rho_2}_{++}$ at fixed $P^{\pi^-}_{lab}$ can be written in the
following form:
\begin{equation}
M^{b_1}_{+-}=ie^{-i\pi\alpha_{b_1}(t)/2}\,\sqrt{-t}\,e^{\Lambda_{b_1}
t/2}\,(s/s_0)^{ \alpha_{b_1}(0)}\,a_
{b_1}\,G_{a_0}(m)\,[2m^2\Gamma_{a_0\eta\pi^0}(m)/\pi]^{1/2}\,,
\end{equation} \begin{equation}
M^{\rho_2}_{++}=e^{-i\pi\alpha_{\rho_2}(t)/2}\,e^{\Lambda_{\rho_2}t/2}
\,(s/s_0)^{ \alpha_{\rho_2}(0)}\,a_{
\rho_2}\,G_{a_0}(m)\,[2m^2\Gamma_{a_0\eta\pi^0}(m)/\pi]^{1/2}\,,
\end{equation}
where $\alpha_j(t)=\alpha_j(0)+\alpha'_j\,t$, $\,a_j$, and
$\Lambda_j/2=\Lambda^0_j/2+\alpha'_j\ln(s/s_0)$ are the
trajectory, residue, and slope of the $j$th Regge pole [one can
accept tentatively $\alpha_{b_1}
(t)\approx-0.21+0.8t/\mbox{GeV}^2$ and
$\alpha_{\rho_2}(t)\approx-0.31+0.8t/\mbox {GeV}^2$],
$\,G_{a_0}(m)=D_{f_0}(m)/[D_{a_0}(m)
D_{f_0}(m)-\Pi^2_{a_0f_0}(m)]$ is the propagator of the mixed
$a^0_0(980)$ resonance [2]. The real situation is rather
interesting. The available data from BNL [3], IHEP [20,21], and
CERN [21] show that, in general, the amplitude $M^{b_1}_{+-}$ is
not required at all to describe the $t$ distributions ($dN/dt$) of
the $\pi^-p\to a_0^0(980)n\to(\eta\pi^0)_S\,n$ reaction events in
the $a_0^0(890)$ mass region. All the data for $0\leq
-t\leq(0.6-0.8)\,$GeV$^2$ are excellently approximated by a
simplest exponential form $C\exp( \Lambda t)$ [4,20,21]
corresponding to the amplitude $M^{\rho_2}_{++}$ nonvanishing for
$t\to0$ [4]. For example, the fit to the normalized BNL data
[3,24] for the differential cross section $d\sigma/dt$ of the
reaction $\pi^-p\to a^0_0(980)n\to( \eta\pi^0)_S\,n$, shown in
Fig. 1(c) by the solid curve, gives $\chi^2/n.d.f.=15.75/22$ and
$d\sigma/dt=[(945.8\pm46.3)\mbox{nb/GeV}^2]\exp[t(4.729
\pm0.217)/\mbox{GeV}^2]$. That is way we consider first of all the
case with the $\rho_2$ and $\pi$ exchanges only.

In Fig. 1(c) the dashed curve shows the differential cross section
due to the $\pi$ exchange, $d\sigma^\pi/dt=\int|M^\pi_{+-}|^2dm $,
corresponding to the region of integration over $m $ from 0.8 to
1.2 GeV at $P^{\pi^-}_{lab}=18.3$ GeV and $
\Lambda_\pi/2\approx4.5$ GeV$^{-2}$ [25,26]. When constructing
this curve for $d\sigma^\pi/dt$, the curve for $|\Pi_{a_0f_0}(m)|$
in Fig. 1(a), as well as the curves in Figs. 1(d)--1(f)
illustrating the expected polarization effect, we used the
following tentative values of the $f_0(980)$ and $a_0(980)$
resonance parameters: $m_{f_0}\approx0.980$ GeV,
$\,g^2_{f_0\pi^+\pi^-}/16\pi\approx\frac{2}{3} \,0.1$\,GeV$^2$,
$\,g^2_{f_0K^+K^-}/16\pi\approx\frac{1}{2}\,0.4$\,GeV$^2$,
$\,\delta_B(m)\approx35.5^\circ+47^\circ m/\mbox{GeV}$,
$\,m_{a_0}\approx0.9847$ GeV, $\,g^2_{a_0K^+K^-}/16\pi\approx
g^2_{f_0K^+K^-}/16\pi\approx\frac{1}{2}\,0.4$\,GeV$^2$, and
$\,g^2_{a_0\eta\pi^0}/16\pi\approx0.25$\,GeV$^2$, in addition, see
Refs. [2,23,27--31]. Note that a strong variation (by about
$90^\circ$) of the phase of the amplitude $\Pi_{a_0f_0}(m)$
between the $K^+K^-$ and $K^0\bar K^0$ thresholds [see Fig. 1(b)
and Eq. (3)], being crucial for polarization phenomena, is
independent of the $f_0(980)$ and $a^0_0(980)$ resonance
parameters. On integrating $d\sigma^\pi/dt$ over $t$, we obtain
$\sigma^\pi\approx10.9$ nb, which makes up about 5.5\% of the
total $\pi^-p\to a^0_0(980)n\to(\eta\pi^0 )_Sn$ reaction cross
section, which is $\approx200$ nb at 18.3 GeV [4,24]. Let us
emphasize that the indicated value of $\sigma^\pi$ should be
considered as its rather reliable lower bound [2,4]. At the
maximum, located near $t\approx-0.0149$ GeV$^2$, $\,d
\sigma^\pi/dt\approx139$ nb/GeV$^2$, which accounts for
approximately 14.7\% of $(d\sigma/dt)|_{t\approx0}$, see Fig.
1(c). However, the main point is that the whole value of
$d\sigma^\pi/dt$ at given $t$, in fact, comes from the narrow
region of $m$ near the $K\bar K$ thresholds, see Fig. 1(a),
whereas the values of the total differential cross section
$d\sigma/dt$ are assembled over the $m$ region which is at least
by an order of magnitude wider. Thus, at low $-t$ and $m$ near the
$K\bar K$ thresholds, the $\pi$ exchange contribution can be quite
comparable with that of the $G$-parity conserving $\rho_2$
exchange. In Figs. 1(d)--1(f) are shown
$d\sigma/dm=\int[|M^{\rho_2}_{++}|^2+|M^\pi_{+-}|^2]dt$,
$\,d\sigma^{\rho_2}/dm=\int|M^{\rho_2}_{++}|^2dt$, $\,I(m)=\int
I(t,m)dt=\int2\mbox{Im}[M^{\rho_2}_{++}(M^\pi_{+-})^*]dt$,
pertaining to the $-t$ region from 0 to 0.025 GeV$^2$ at
$P^{\pi^-}_{lab}=18.3$ GeV, and the corresponding asymmetry
$A(0\leq-t\leq0.025\mbox{\,GeV}^2,\,m)$. In so doing, the
parameters of the $\rho_2$ exchange, which we substitute in Eq.
(5), correspond to the above-mentioned fit to the BNL data; see
Fig. 1(c). Notice that the $I(m)$ and asymmetry are determined
only up to the sign because the relative sign of the $\rho_2$ and
$\pi$ exchanges is unknown. Figures 1(d)--1(f) show that the
polarization effect caused by the interference between the
amplitudes $M_{++}^{\rho_2}$ and $M_{+-}^\pi$ is highly
significant. A natural measure of the effect is the magnitude of a
distinctive jump of the asymmetry, which takes place in the $m$
region from 0.965 to 1.01 GeV. As is seen from Fig. 1(f) the
corresponding difference between the maximal and minimal values of
the asymmetry smoothed at the expense of the finite $\eta\pi^0$
mass resolution turns out to be approximately equal to 0.95 in
this mass region, see the dotted curve in Fig. 1(f) and the figure
caption. Note that any noticeable variation of the interference
pattern does not arise if one refits the BNL data in Fig. 1(c) by
adding the $\pi$ exchange contribution, indicated in the same
figure, to the $\rho_2$ exchange one. To demonstrate the
polarization effect, we have restricted ourselves to a single
interval of $-t$ from 0 to 0.025 GeV$^2$ only from brevity
considerations. We constructed the figures, analogous to Figs.
1(d)--1(f), for the intervals
$0\leq-t\leq0.05,\,0.1,\,0.2\mbox{\,GeV}^2$ and made sure that the
relative magnitude of the polarization effect is practically
unchanged. Furthermore, we examined the effects of the $b_1$
exchange contribution in detail. We enabled the $b_1$ exchange
contribution to reach 40\% of the total cross section. But even
so, the influence of the $b_1$ exchange is inessential in the low
$-t$ region. The resulting conclusion is that in this case the
smoothed asymmetry pertaining to any interval of
$0\leq-t\leq0.025,...,0.1$ GeV$^2$ must also undergo a jump of
order one in the region $0.965\leq m\leq1.01$ GeV owing to the
$\pi$ exchange admixture.

Finally, we emphasize that observing the asymmetry jump does not
require at all any very high-quality $\eta\pi^0$ mass resolution
that would be absolutely necessary to recognize the
$a^0_0(980)-f_0(980)$ mixing manifestation in the $\eta\pi^0$ mass
spectrum in the unpolarized experiment.

Currently, experimental investigations utilizing the polarized
beams and targets are on the rise. Therefore, this proposal seems
to be quite opportune. The indicated polarization effect can be
investigated at any high energy, for example, in the range from 8
to 100 GeV because of nearness of the $\pi$, $\rho_2$, and $b_1$
Regge trajectories. The relevant experiments on the reaction
$\pi^-p\to\eta\pi^0n$ on a polarized proton target, in principle,
can be realized at KEK, BNL, IHEP, CERN (COMPASS), FNAL, ITEP, and
Institut f\"ur Kernphysik in J\"ulich. Discovery of the
$a^0_0(980)-f_0(980)$ mixing would open one more interesting page
in investigation of the nature of the puzzling $a^0_0(980)$ and
$f_0(980)$ states. Of course, the general idea using polarization
phenomena as an effective tool for the observation of the
$a_0^0(980)-f_0(980)$ mixing connected with the great variation
(by about 90$^\circ$) of the phase of the $a^0_0(980)-f_0(980)$
mixing amplitude in the narrow energy region (8 MeV) between the
$K^+K^-$ and $K^0\bar K^0$ thresholds is also applicable to other
reactions. A more extended discussion of the questions touched on
here will be presented elsewhere.

This work was supported in part by the RFBR Grant No. 02-02-16061
and the Presidential Grant No. 2339.2003.2 for support of Leading
Scientific Schools.

\begin{figure}\centerline{\epsfysize=8.75in\epsfbox{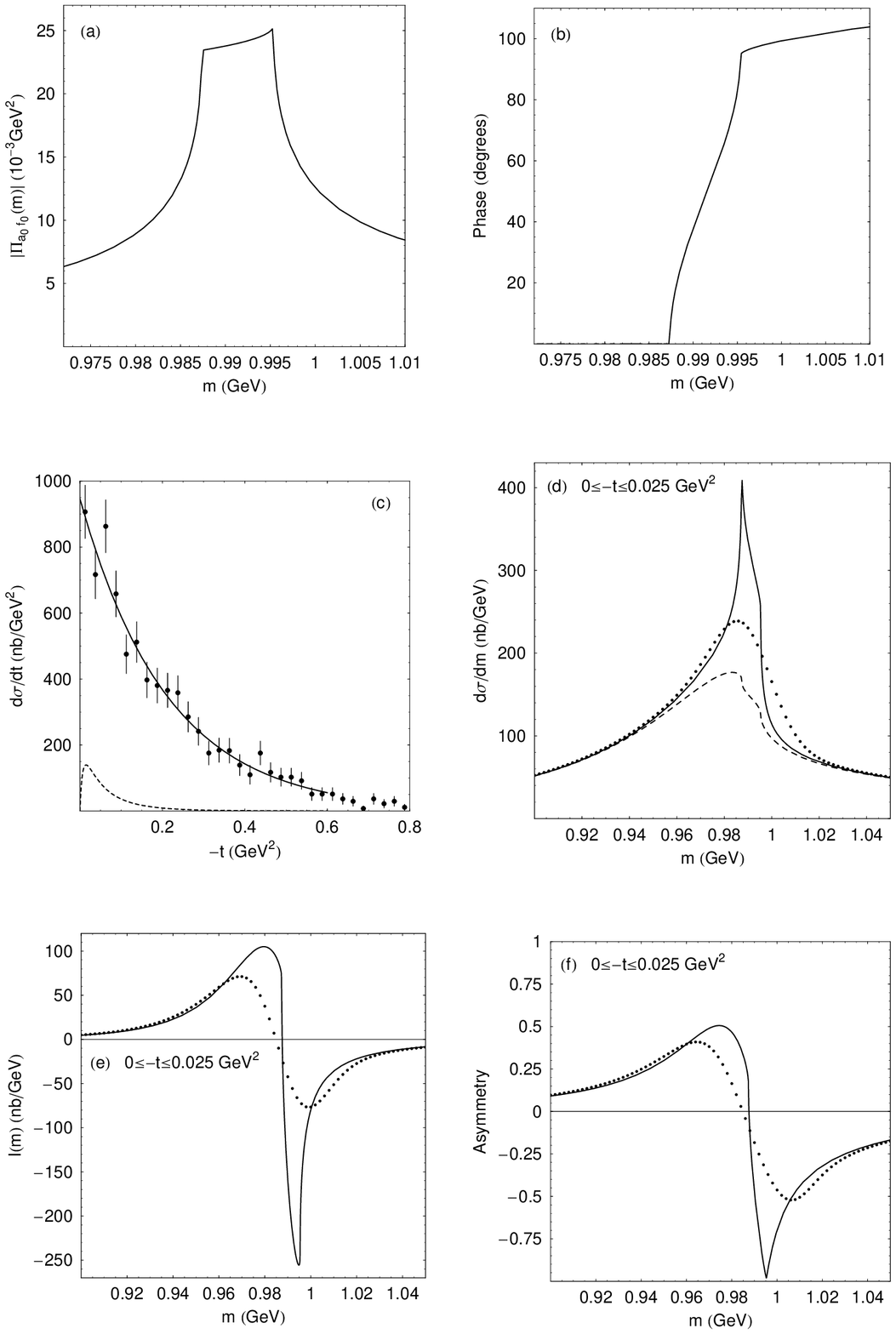}}
\caption{ (a) The modulus of the $a^0_0(980)-f_0(980)$ transition
amplitude, see Eq. (3). (b) The phase of the $a^0_0(980)-f_0(980)$
transition amplitude. (c) The experimental points are the
normalized BNL data for $d\sigma/dt$ of the reaction $\pi^-p\to
a^0_0(980)n\to( \eta\pi^0)_S\,n$ [3,24], the solid curve shows the
fit to the data in the $\rho_2$ exchange model, the dashed curve
shows $d\sigma^\pi/dt$ for the process $\pi^-p\to f_0(980)n\to
a^0_0(980)n\to(\eta\pi^0)_S\,n$ due to the $\pi$ exchange
mechanism only at $P^{\pi^-}_{lab}= 18.3$ GeV. (d),(e),(f) The
manifestation of the $a^0_0(980)-f_0(980)$ mixing in the reaction
$\pi^-p\to a^0_0 (980)n\to (\eta\pi^0)_S\,n$ on a polarized target
at $P^{\pi^-}_{lab}=18.3$ GeV in the $\rho_2$ and $\pi$ exchange
model. The solid curves in (d),(e),(f) show $d\sigma/dm$,
$\,I(m)$, for the region $0\leq-t\leq0.025$ GeV$^2$, and the
corresponding asymmetry $A(0\leq-t\leq0.025\mbox{\,GeV}^2,\,m)$
[17], respectively; the common sign of the $I(m)$ and asymmetry
was chosen arbitrarily. The dashed curve in (d) shows the $\rho_2$
exchange contribution to $d\sigma/dm$. The dotted curves in
(d),(e),(f) show $d\sigma/dm$, $I(m)$, smoothed with the Gaussian
mass distribution with the dispersion of 10 MeV, and the
corresponding asymmetry, respectively.}\end{figure}


\begin{references}
\bibitem{} The asymmetry pertaining to some interval of $t$, $\,A(t_1\leq t\leq
           t_2,m)$, or to some interval of $m$, $\,A(t,m_1\leq m\leq m_2)$,
           is defined by the ratio of the corresponding integrals of $I(t,m)$ and
           $d^2\sigma /dtdm$ over $t$ or over $m$.
\bibitem{} The process $\pi^-p\to f_0(908)n\to a^0_0(980)n\to(\eta\pi^0)_S\,n$
           can also occur via the $a_1$ exchange. The estimates show, however,
           that the contribution of the amplitude $M^{a_1}_{++}$ can be
           neglected in comparison with the other ones.
\bibitem{} The experimental points in Fig. 1(c) correspond to the BNL data
           for $dN/dt$ at $P^{\pi^-}_{lab}=18.3$ GeV [3] normalized on the
           $a^0_2(1320)$ formation cross section in the reaction $\pi^-p\to
           a_2^0(1320)n$ in such a way as it was done in Ref. [4].
           The total cross section $\sigma$ for
           the reaction $\pi^-p\to a^0_0(980)n\to(\eta\pi^0 )_Sn$ at 18.3
           GeV is approximately equal to 200\,nb [4]. This number we refer to
           the $m$ region from 0.8 to 1.2 GeV and to the whole region of $t\leq0$.
           Note that the above value of $\sigma$ is in close agreement with the
           estimate presented in Ref. [20].
\end{references}
\end{document}